# Student difficulties with representations of quantum operators corresponding to observables

Emily Marshman and Chandralekha Singh

*University of Pittsburgh, Department of Physics and Astronomy, 3941 O'Hara St., Pittsburgh, PA 15260*

**Abstract.** Dirac notation is used commonly in quantum mechanics. However, many upper-level undergraduate and graduate students in physics have difficulties with representations of quantum operators corresponding to observables especially when using Dirac notation. To investigate these difficulties, we administered free-response and multiple-choice questions and conducted individual interviews with students in advanced quantum mechanics courses. We discuss the analysis of data on the common difficulties found.

## I. INTRODUCTION

Learning quantum mechanics (QM) is challenging even for advanced undergraduate and graduate students [1-4]. Investigations of student difficulties in QM are important for developing curricula and pedagogies that help students develop a solid grasp of QM [5-8]. However, no prior research studies have focused on student difficulties with the representations of quantum operators corresponding to observables in the context of Dirac notation, a compact and convenient notation commonly used in advanced QM.

Here we discuss an investigation of student difficulties with representations of quantum operators corresponding to observables acting on quantum states in the position and momentum representations when using Dirac notation. The ability to express a concept in different representations is a hallmark of expertise. Physics experts use multiple representations of concepts and have the ability to construct, interpret, and transform between different representations of knowledge. In many situations, the use of concrete representations in QM can facilitate better understanding of abstract concepts. Therefore, one goal of many advanced QM courses is to help students learn how to represent a concept in different representations and use a representation that makes the problem solving task easier in a given context.

## II. BACKGROUND

We first discuss the requisite knowledge needed to understand representations of quantum operators corresponding to observables acting on quantum states, shown in Table I. Quantum states, which contain all information about the state of a quantum system and hermitian operators, which correspond to physical observables, can be expressed in a variety of representations. A position eigenstate with eigenvalue $x'$ in Dirac notation is written as $|x'\rangle$. Ignoring the normalization issue related to position and momentum eigenfunctions, in the position representation, this position eigenstate $|x'\rangle$ can be represented as $\langle x|x'\rangle = \delta(x - x')$, which is a highly localized wavefunction about $x = x'$. The virtue of representing any quantum state in the position representation is that one can visualize how localized a wavefunction is in real space. On the other hand, the same position eigenstate, in the momentum representation, can be represented as $\langle p|x'\rangle = e^{-ipx'/\hbar}$, which is a completely delocalized wavefunction. Similarly, a momentum eigenstate with eigenvalue $p'$ in Dirac notation is written as $|p'\rangle$. In the position representation, this momentum eigenstate $|p'\rangle$ can be represented as $\langle x|p'\rangle = e^{ip'x/\hbar}$ (which is a delocalized wavefunction in real space). On the other hand, in the momentum representation, the same momentum eigenstate $|p'\rangle$ can be represented as $\langle p|p'\rangle = \delta(p - p')$, which is a highly localized wavefunction.

**Table I.** Operators acting on quantum states in Dirac notation (DN), position representation (PR), and momentum representation (MR), ignoring normalization issues.

| | |
|---|---|
| DN | Position operator $\hat{x}$ acting on position eigenstate $\|x'\rangle$: $\hat{x}\|x'\rangle = x'\|x'\rangle$ |
| PR | $\langle x\|\hat{x}\|x'\rangle = x'\langle x\|x'\rangle = x'\delta(x - x') = x\delta(x - x')$ |
| MR | $\langle p\|\hat{x}\|x'\rangle = i\hbar\, \partial/\partial p\, \langle p\|x'\rangle = i\hbar\, \partial/\partial p\, e^{-ipx'/\hbar} = x' e^{-ipx'/\hbar}$ |
| DN | Momentum operator $\hat{p}$ acting on momentum eigenstate $\|p'\rangle$: $\hat{p}\|p'\rangle = p'\|p'\rangle$ |
| PR | $\langle x\|\hat{p}\|p'\rangle = -i\hbar\, \partial/\partial x\, \langle x\|p'\rangle = -i\hbar\, \partial/\partial x\, e^{ip'x/\hbar} = p' e^{ip'x/\hbar}$ |
| MR | $\langle p\|\hat{p}\|p'\rangle = p'\langle p\|p'\rangle = p'\delta(p - p') = p\delta(p - p')$ |
| DN | Generic operator $\hat{Q}$ acting on a generic quantum state $\|\Psi\rangle$: $\hat{Q}\|\Psi\rangle$ |
| PR | $\langle x\|\hat{Q}\|\Psi\rangle = Q(x, -i\hbar\, \partial/\partial x)\Psi(x)$ |
| MR | $\langle p\|\hat{Q}\|\Psi\rangle = Q(i\hbar\, \partial/\partial p, p)\Psi(p)$ |

Furthermore, in the position representation, the position operator $\hat{x}$ can be represented as $x$ (which is simply a multiplication by $x$) and in the momentum representation, the same position operator can be represented as $i\hbar\, \partial/\partial p$. Also, in the position representation, the momentum operator $\hat{p}$ can be represented as $-i\hbar\, \partial/\partial x$ and in the momentum representation, the same momentum operator can be represented as $p$ (which is simply a multiplication by $p$).

The position operator $\hat{x}$ acting on a position eigenstate $|x'\rangle$ with eigenvalue $x'$ yields the following eigenvalue equation: $\hat{x}|x'\rangle = x'|x'\rangle$. To represent $\hat{x}|x'\rangle$ in the position representation, one must project $\hat{x}|x'\rangle$ onto the position eigenstates $|x\rangle$, i.e., $\langle x|\hat{x}|x'\rangle$. This expression can be written as $\langle x|\hat{x}|x'\rangle = x\delta(x - x')$. Also, $x\delta(x - x') = x'\delta(x - x')$ since the Dirac delta function $\delta(x - x')$, which corresponds to a quantum state in which the position of the particle is well defined, is zero for all positions except when $x = x'$. To represent $\hat{x}|x'\rangle$ in the momentum representation, one must project $\hat{x}|x'\rangle$ onto the momentum eigenstates $|p\rangle$, i.e., $\langle p|\hat{x}|x'\rangle$. This expression can be written as $\langle p|\hat{x}|x'\rangle = i\hbar\, \partial/\partial p\, \langle p|x'\rangle = i\hbar\, \partial/\partial p\, e^{-ipx'/\hbar} = x' e^{-ipx'/\hbar}$. Similarly, a momentum operator $\hat{p}$ acting on a momentum eigenstate $|p'\rangle$ with eigenvalue $p'$ yields the following eigenvalue equation: $\hat{p}|p'\rangle = p'|p'\rangle$. To represent $\hat{p}|p'\rangle$ in the position representation, one must project $\hat{p}|p'\rangle$ onto the position eigenstates $|x\rangle$, i.e., $\langle x|\hat{p}|p'\rangle$. This expression can be written as $\langle x|\hat{p}|p'\rangle = -i\hbar\, \partial/\partial x\, \langle x|p'\rangle = -i\hbar\, \partial/\partial x\, e^{ip'x/\hbar} = p' e^{ip'x/\hbar}$. To represent $\hat{p}|p'\rangle$ in the momentum representation, one must project $\hat{p}|p'\rangle$ onto the momentum eigenstates $|p\rangle$, i.e., $\langle p|\hat{p}|p'\rangle$. This expression can be written as $\langle p|\hat{p}|p'\rangle = p\delta(p-p') = p'\delta(p-p')$. Table I summarizes different representations of the position operator acting on a position eigenstate, momentum operator acting on a momentum eigenstate, and a generic quantum operator $\hat{Q}$ corresponding to a physical observable acting on a generic quantum states.

## III. METHODOLOGY

Student difficulties with recognizing and transforming operators corresponding to observables acting on quantum states from one representation to another were investigated by administering open-ended and multiple-choice questions after traditional instruction in relevant concepts to upper-level undergraduate (UG) and graduate (G) students and observing difficulties on in-class quizzes. The UG students were enrolled in a junior/senior level QM course and the G students were enrolled in a first-year core graduate QM course. See Table II for a list of the questions that were administered to students. The multiple-choice questions were administered to 184 upper-level UG students at four U.S. universities (see Table II, questions Q1 and Q5). The open-ended quiz questions were administered to 62 UG and 68 G students at the University of Pittsburgh (see Table II, questions Q2-Q4 and Q6) in which students had to transform the given expression to a representation other than the one in which it was given. We hypothesized that while the representational transformation task may be easy for those who have a conceptual understanding of what those expressions mean in QM, the task may be difficult for those students who had not learned to make sense of the expressions conceptually.

Student difficulties were also investigated by conducting individual interviews with a total of 23 UG and G student volunteers enrolled in QM courses. The individual interviews employed a think-aloud protocol to better understand the rationale for student written responses. During the semi-structured interviews, students were asked to "think aloud" while answering the questions. Students first read the questions on their own and answered them without interruptions except that they were prompted to think aloud if they were quiet for a long time. After students had finished answering a particular question to the best of their ability, we asked them to further clarify and elaborate issues that they had not clearly addressed earlier.

**Table II.** Questions related to momentum operator $\hat{p}$ and a generic operator $\hat{Q}$ acting on quantum states in position and momentum representation and number of students (N) who answered each question. The correct answer is bolded.

| Questions | N |
|---|---|
| **Q1**. $|p'\rangle$ is the momentum eigenstate with eigenvalue $p'$ for a particle confined in one spatial dimension. Choose all of the following statements that are correct.<br>1. $\langle p|\hat{p}|p'\rangle = p'\langle p|p'\rangle = p'\delta(p-p')$<br>2. $\langle x|\hat{p}|p'\rangle = p'\langle x|p'\rangle = p' e^{ip'x/\hbar}$<br>3. $\langle x|\hat{p}|p'\rangle = -i\hbar\, \partial/\partial x\, \langle x|p'\rangle = -i\hbar\, \partial/\partial x\, e^{ip'x/\hbar}$<br>**A. all of the above**, B. 1 only, C. 1 and 2 only<br>D. 1 and 3 only, E. 2 and 3 only | 184 UG |
| **Q2.** Fill in the blank (without using Dirac notation): $\langle x|\hat{p}|p'\rangle =$ _______ | 62 UG |
| | 68 G |
| **Q3.** Fill in the blank (without using Dirac notation): $\langle p|\hat{p}|p'\rangle =$ _______ | 62 UG |
| | 68 G |
| **Q4.** Fill in the blank $\hat{x}\delta(x-x') =$ _______ | 62 UG |
| | 68 G |
| **Q5**. An operator $\hat{Q}$ corresponding to a physical observable in the position representation can be expressed as $Q(x, -i\hbar\, \partial/\partial x)$. Choose all of the following statements that are correct.<br>1. $\hat{Q}|x\rangle = Q(x, -i\hbar\, \partial/\partial x)$<br>2. $\langle x|\hat{Q}|\Psi\rangle = Q(x, -i\hbar\, \partial/\partial x)\Psi(x)$<br>3. $\langle x|\hat{Q}|\Psi\rangle = \langle \Psi|\hat{Q}|x\rangle$<br>A. All of the above, **B. 2 only,** C. 1 and 2 only<br>D. 1 and 3 only, E. 2 and 3 only | 184 UG |
| **Q6.** Fill in the blank (without using Dirac notation): $\langle x|\hat{Q}|\Psi\rangle =$ _______ | 62 UG |
| | 68 G |

## IV. FINDINGS

### A. Results: Undergraduate Student Difficulties

**Inconsistent responses when expressing $\langle x|\hat{p}|p'\rangle$ without using Dirac notation.** Table III shows that, in response to question Q1, only 23% of the students provided the correct response option A. This low percentage is due to the fact that many students did not realize that all three statements listed in question Q1 are true. Table III also shows that the most popular incorrect option selected for question Q1 was option C (statements 1 and 2 only). Interviews suggest that many students selected options 1 and 2 as correct because they were able to act on $|p'\rangle$ with the momentum operator and pull the eigenvalue $p'$ out of the bracket, i.e., $\langle p|\hat{p}|p'\rangle = p'\langle p|p'\rangle$ or $\langle x|\hat{p}|p'\rangle = p'\langle x|p'\rangle$. Then, they were able to reason that $\langle p|\hat{p}|p'\rangle = p'\langle p|p'\rangle = p'\delta(p-p')$ or $\langle x|\hat{p}|p'\rangle = p'\langle x|p'\rangle = p' e^{ip'x/\hbar}$. However, many students did not realize that statement 3 is also a correct way to transform $\langle x|\hat{p}|p'\rangle$ into a form not involving the Dirac notation. In statement 3, one can check the validity of the statement by realizing that the statement involves a representation of the momentum operator acting on a momentum eigenstate in the position representation. Therefore, to express it without using Dirac notation, one can represent the momentum operator $\hat{p}$ in the position representation as $-i\hbar\,\partial/\partial x$ and the momentum eigenstate $|p'\rangle$ in the position representation as $\langle x|p'\rangle = e^{ip'x/\hbar}$.

Table IV also shows that, on question Q1, 45% of the students selected an option that did not include both statements 2 and 3 (i.e., options C or D). This dichotomy indicates that many students did not realize that the two methods (used in statements 2 and 3) for evaluating the expression $\langle x|\hat{p}|p'\rangle$ are equivalent ways to find an expression for $\langle x|\hat{p}|p'\rangle$ without using the Dirac notation. In interviews, many students had difficulty recognizing whether $\langle x|\hat{p}|p'\rangle = -i\hbar\,\partial/\partial x\,\langle x|p'\rangle$ is true or not mainly because they did not realize that $\langle x|\hat{p}|p'\rangle$ is conceptually equivalent to the momentum operator $\hat{p}$ acting on $|p'\rangle$ (i.e., $\hat{p}|p'\rangle$) expressed in the position representation.

In interviews, students were asked to transform $\hat{p}$ or $|p'\rangle$ to the position representation. Many of them were separately able to write $\hat{p}$ as $-i\hbar\,\partial/\partial x$ and $|p'\rangle$ as $e^{ip'x/\hbar}$ in the position representation. However, students often struggled to write $\langle x|\hat{p}|p'\rangle$ without using Dirac notation. From an expert point of view, representing $\hat{p}|p'\rangle$ in the position representation is the same as expressing it as $\langle x|\hat{p}|p'\rangle$. However, many interviewed students did not know that the expression $\langle x|\hat{p}|p'\rangle$ is equivalent to writing $\hat{p}|p'\rangle$ in the position representation (i.e., projecting $\hat{p}|p'\rangle$ onto the position eigenstates $|x\rangle$).

In their responses to the open-ended questions, students sometimes displayed specific difficulties when asked to write $\langle x|\hat{p}|p'\rangle$, $\langle p|\hat{p}|p'\rangle$, $\hat{x}\delta(x-x')$, and $\langle x|\hat{Q}|\Psi\rangle$ in a form other than the one in which the expression was given. Below, we discuss some of the difficulties identified in this context.

**Invoking an incorrect orthogonality condition.** Table III shows that, in response to question Q2, only 15% of the UG students were able to write $\langle x|\hat{p}|p'\rangle$ in a form not involving Dirac notation, i.e., $p'e^{ip'x/\hbar}$. Students were not penalized if they wrote an incorrect sign in the exponent. Table IV shows that 5% of the students stated that $\langle x|\hat{p}|p'\rangle = 0$. Interviews suggest that this difficulty often stems from the fact that that students incorrectly invoked an orthogonality condition between momentum and position eigenstates, i.e., $\langle x|p'\rangle = 0$. A similar difficulty has been found in the context of spin in that many students incorrectly think that an eigenstate of one component of spin is orthogonal to an eigenstate of another component of spin [5].

**Invoking an incorrect normalization condition.** Table III shows that, in response to question Q3, only 27% of the UG students wrote the correct expression, $\langle p|\hat{p}|p'\rangle = p'\delta(p-p')$. Table IV also shows that 8% of the students wrote $\langle p|\hat{p}|p'\rangle = p'$ or ignored the eigenvalue $p'$ and wrote $\langle p|\hat{p}|p'\rangle = 1$. Other students incorrectly wrote that $\langle p|\hat{p}|p'\rangle = 0$. Interviews suggest that these difficulties often stem from incorrectly assuming that $\langle p|p'\rangle = 1$ or $\langle p|p'\rangle = 0$. In interviews, students often incorrectly claimed that $\langle p|p'\rangle = 1$ if $p = p'$. Interviews suggest that this type of difficulty was often the result of confusing the Kronecker delta and the Dirac delta function.

**Table III.** Percentage of students correctly answering questions related to momentum operator $\hat{p}$ and a generic operator $\hat{Q}$ in position and momentum representations. Correct responses are in bold.

| Question | Percentages of responses |
|---|---|
| Q1 | **A (23%)** B (17%) C (30%) D (15%) E (14%) |
| Q2 | **15% of UG, 69% of G** |
| Q3 | **27% of UG, 84% of G** |
| Q4 | **26% of UG, 66% of G** |
| Q5 | A (15%) **B (35%)** C (24%) D (11%) E (13%) |
| Q6 | **6% of UG, 31% of G** |

**Incorrectly stating $\hat{x}\delta(x-x') = x'$.** Table III shows that, in response to Q4, 26% of the UG students were able to write another correct expression for $\hat{x}\delta(x-x')$. Responses were considered correct if they were of the form $x\delta(x-x')$, $x'\delta(x-x')$, $x\langle x|x'\rangle$, $\langle x|x|x'\rangle$, or $\langle x|\hat{x}|x'\rangle$. Table IV shows that in response to Q4, 23% of the students incorrectly wrote $\hat{x}\delta(x-x') = x'$. Interviews suggest that students often claimed that $\hat{x}\delta(x-x') = x'$ due to mathematical and/or conceptual difficulties. For example, some students incorrectly argued mathematically that the Dirac delta function picks out the value $x'$ even when no integration is involved or that $\hat{x}\delta(x-x')$ is the position operator $\hat{x}$ acting on its eigenstate $\delta(x-x')$ and so it would give the eigenvalue $x'$. Others made conceptual arguments claiming that, in QM, the position operator acting on a position eigenfunction corresponds to a measurement of position and must yield a position eigenvalue $x'$ according to the postulates of QM. A similar difficulty has been found in the context of the Hamiltonian operator $\hat{H}$ acting on an energy eigenstate $|\psi_n\rangle$—many students incorrectly claim that $\hat{H}|\psi_n\rangle = E_n$ [5] because $\hat{H}$ acting on its eigenstate corresponds to the measurement of energy. These students omitted writing the state $|\psi_n\rangle$ on the right hand of the eigenvalue equation, i.e., $\hat{H}|\psi_n\rangle = E_n|\psi_n\rangle$, similar to students who incorrectly wrote $\hat{x}\delta(x - x') = x'$ and did not realize that $\hat{x}\delta(x - x') = x'\delta(x - x')$.

**Table IV.** Percentages of UG students displaying difficulties with position and momentum operators in position or momentum representation.

| Questions and difficulties | % |
|---|---|
| **Q1**: Inconsistent responses, e.g., claiming that $\langle x|\hat{p}|p'\rangle = p'\langle x|p'\rangle = p'e^{ip'x/\hbar}$ is correct but $\langle x|\hat{p}|p'\rangle = -i\hbar\,\partial/\partial x\,\langle x|p'\rangle = -i\hbar\,\partial/\partial x\,e^{ip'x/\hbar}$ is not correct | 45% |
| **Q2:** Invoking an incorrect orthogonality condition between position and momentum eigenstates, e.g., $\langle x|\hat{p}|p'\rangle = p'\langle x|p'\rangle = 0$ | 5% |
| **Q3**: Invoking an incorrect normalization condition, e.g., $\langle p|\hat{p}|p'\rangle = p'\langle p|p'\rangle = p'$ or $\langle p|\hat{p}|p'\rangle = p'\langle p|p'\rangle = 1$ if $p = p'$ | 8% |
| **Q4**: Incorrectly stating that $\hat{x}\delta(x-x') = x'$ | 23% |

**Difficulty in representing a generic operator acting on a generic quantum state.** Table III shows that, in response to question Q5, only 35% of the UG students correctly identified that $\langle x|\hat{Q}|\Psi\rangle = Q(x, -i\hbar\,\partial/\partial x)\Psi(x)$ and in Q6, only 6% correctly wrote $\langle x|\hat{Q}|\Psi\rangle = Q(x, -i\hbar\,\partial/\partial x)\Psi(x)$. Responses were considered correct if students wrote both the operator $\hat{Q}$ and state $|\Psi\rangle$ in position representation. In a prior research study [9], over 80% of the UG students were able to recall that $\langle x|\Psi\rangle$ is the wave function in position representation. Interviews suggest that even in the present study, some students knew that the state $|\Psi\rangle$ in the position representation is $\langle x|\Psi\rangle = \Psi(x)$ and also that $\hat{Q}|\Psi\rangle$ can be represented as $Q(x, -i\hbar\,\partial/\partial x)\Psi(x)$ in the position representation. However, they struggled with the fact that $\langle x|\hat{Q}|\Psi\rangle$ is conceptually equivalent to expressing the operator $\hat{Q}$ acting on state $|\Psi\rangle$ in position representation.

### B. Results: Graduate Student Difficulties

Table III shows that, in response to questions Q2 and Q3, 69% of the G students were able to write $\langle x|\hat{p}|p'\rangle$ and 84% were able to write $\langle p|\hat{p}|p'\rangle$ without using Dirac notation. Interviews and written responses indicate that G students were facile at transforming $\langle x|\hat{p}|p'\rangle$ and $\langle p|\hat{p}|p'\rangle$ to a form not involving Dirac notation because many had committed to memory the expressions for $\langle x|p'\rangle$ and $\langle p|p'\rangle$. For example, G students usually found the expression for $\langle x|\hat{p}|p'\rangle$ by acting on $|p'\rangle$ with the momentum operator $\hat{p}$, pulling the eigenvalue $p'$ out of the bracket, and writing the momentum eigenstate in position representation (i.e., $\langle x|\hat{p}|p'\rangle = p'\langle x|p'\rangle = p'e^{ip'x/\hbar}$). However, in interviews, when asked about the interpretation of $\langle x|\hat{p}|p'\rangle$, many of them did not know that $\langle x|\hat{p}|p'\rangle$ is equivalent to representing $\hat{p}|p'\rangle$ in the position representation.

**Difficulty with a generic operator acting on a generic quantum state.** Table III shows that, in response to question Q6, only 31% of the G students realized that $\langle x|\hat{Q}|\Psi\rangle = \hat{Q}(x, -i\hbar\,\partial/\partial x)\Psi(x)$. In written responses and interviews, G students attempted unproductive approaches to transform the expression into a different representation (not involving Dirac notation) such as inserting the identity operator in terms of a complete set of position eigenstates or inserting the identity operator in terms of a complete set of eigenstates of $\hat{Q}$. In interviews, students were asked to explain in words their interpretation of $\langle x|\hat{Q}|\Psi\rangle$. Many students were unable to articulate that the expression $\langle x|\hat{Q}|\Psi\rangle$ is equivalent to writing $\hat{Q}$ acting on a generic quantum state $|\Psi\rangle$ (i.e., $\hat{Q}|\Psi\rangle$) in the position representation (i.e., projecting $\hat{Q}|\Psi\rangle$ onto the position eigenstates $|x\rangle$).

## V. SUMMARY

Helping advanced students become facile with different representations of the concepts discussed here requires that students integrate mathematical and conceptual aspects of the QM formalism involving Dirac notation. We find that the UG students have many common difficulties in transforming quantum operators corresponding to observables acting on position or momentum eigenstates from one form to another after traditional instruction in relevant concepts. Interviews and written responses indicate that they often knew how to represent, e.g., $\hat{p}$ and $|p'\rangle$ in the position representation but they did not realize that $\langle x|\hat{p}|p'\rangle$ is equivalent to representing $\hat{p}$ acting on a momentum eigenstate $|p'\rangle$ in the position representation. G students were more facile than UG students at these representational tasks but interviews suggest that for many of them, this facility was mainly due to the fact that they had memorized expressions for $\langle x|p'\rangle$ and $\langle p|p'\rangle$ and were following procedures. In other words, they were often unable to explain in words, e.g., that the expression $\langle x|\hat{p}|p'\rangle$ is equivalent to $\hat{p}|p'\rangle$ in the position representation.

Both UG and G students struggled in translating $\langle x|\hat{Q}|\Psi\rangle$ into a notation not involving Dirac notation. It is interesting to note that G students often attempted complicated unproductive strategies to perform this transformation. Interviews suggest that many students did not understand that $\langle x|\hat{Q}|\Psi\rangle$ is equivalent to $\hat{Q}|\Psi\rangle$ in the position representation. In particular, many students had difficulty transforming $\langle x|\hat{Q}|\Psi\rangle$ to a form not involving Dirac notation despite knowing how to represent $\hat{Q}|\Psi\rangle$ in the position representation. These findings can be used as resources for developing curricula and pedagogies to help advanced students in QM courses develop facility with QM concepts in different representations and be able to transform quantum operators acting on various quantum states from one representation to another when using Dirac notation.

### ACKNOWLEDGEMENTS

We thank the NSF for award PHY-1505460.

[1] C. Singh, Am. J. Phys. **69**, 885 (2001); **76**, 277 (2008); **76**, 400 (2008); arXiv:1602.05655; arXiv:1602.05664.
[2] M. Wittmann et al., Am. J. Phys. **70**, 218 (2002).
[3] D. Zollman et al., Am. J. Phys. **70**, 252 (2002).
[4] G. Zhu and C. Singh, Am. J. Phys. **79**, 499 (2011); **80**, 252 (2012); PRST PER **8**, 010117 (2012); PRST PER **8**, 010118 (2012); **9**, 010101 (2013); arXiv:1602.05619.
[5] C. Singh and E. Marshman, PRST PER **11**, 020117 (2015); arXiv:1509.07740; arXiv:1603.02948.
[6] B. Brown et al. PRST PER, **12**, 010121 (2016).
[7] G. Passante et al. PRST PER **11**, 020111 (2015).
[8] E. Marshman and C. Singh, PRST PER **11**, 020119 (2015); Eur. J. Phys. **37**, 024001 (2016).
[9] arXiv:1509.04084; arXiv:1510.01296; arXiv:1510.01319.